\documentclass[AMA,STIX1COL]{WileyNJD-v2}
\articletype{Article Type}

\usepackage{amsmath}
\usepackage{amssymb}
\usepackage{subfigure}
\usepackage{color}
\usepackage{epsfig}

\begin{document}
	\newcommand{\sign}[1]{\mathrm{sgn}(#1)} 
	\title{Adaptive Event-triggered Control For Strict-feedback Systems With Time-varying Parameters}
 
	\author[1]{Yan Tan}
	\authormark{Tan \textsc{et al}}
	\author[1]{Liucang Wu*}
	\author[2]{Wenqi Liu}
	
	\address[1]{\orgdiv{Faculty of Science}, \orgname{Kunming University of Science and Technology}, \orgaddress{\state{Kunming}, \country{China}}} 
	\address[2]{\orgdiv{Data Science Research Center, Faculty of Science}, \orgname{Kunming University of Science and Technology}, \orgaddress{\state{Kunming}, \country{China}}} 
		
	\corres{Yan Tan, Faculty of Science, Kunming University of Science and Technology, Kunming, 650500, China.\\Email: tanyan@stu.kust.edu.cn. \\ *Liucang Wu, Faculty of Science, Kunming University of Science and Technology, Kunming, 650500, China.\\Email: wuliucang@163.com.}
	\fundingInfo{National Natural Science Foundation of China under grant (No.12261051).}
	
\abstract[Summary]{
	In this article, we develop a new adaptive event-triggered asymptotic control scheme for strict-feedback systems with fast time-varying parameters. To deal with time-varying parameters with unknown variation boundaries in the feedback path and the input path, we construct three adaptive laws for parameter estimation, two for the uncertain parameters in the feedback path and one for the uncertain parameters in the input path. In particular, two sets of tuning functions are introduced to avoid over-parametrization. Additionally, an event-triggering mechanism is embedded in this adaptive control framework to reduce the data transmission from the controller to the actuator. We also introduce a soft sign function to handle the perturbations caused by sampling errors to achieve asymptotic stability and avoid the so-called parameter drift. The stability analysis shows that the closed-loop system is globally uniformly asymptotically stable and the Zeno behavior can be excluded. Simulation results verify the effectiveness and performance of the proposed adaptive scheme.
}

	
	\maketitle
	
\section{Introduction} 
Unlike traditional periodic time-triggered control, event-triggered control only updates the control signal when the triggering conditions are violated, which indicates that communication and computational resources can be used more efficiently \cite{2022-zhangzhirong-auto,ZQX2019,ZGP2021,2022-zhangzhirong-tac,ZhangXM2023}. Due to these advantages, the event-triggered control has garnered significant attention in recent years, leading to the successful development of numerous event-triggering algorithms \cite{QWH2023, QWH2022, QWH2021, QWH2022-1, WPMH2013, WY2022, XieYK2023}, and its application to various real-world problems has also been observed \cite{Kaze2022, GeXH2022, GuZ2023}.

It is well-established that adaptive control, endowed with online estimation/learning capabilities through identification/compensation mechanisms, is highly effective in dealing with unknown parameters and improving system performance \cite{LiKun2023,Yantan2023,ZhaoK2016,ZhaoK2022}. Existing research has shown that adaptive event-triggered control can improve the transient performance of closed-loop systems while reducing data transmission by setting suitable triggering conditions \cite{Xing2017,XLT20174,HuangYX2019,LongJ2022,XLT20172,LD2023,XLT20173,KarafyllisI2018,Huang2020}. Based on the different triggering objects, these existing results are generally divided into two categories: signal transmission triggering \cite{Xing2017,XLT20172,XLT20173} and parameter estimation triggering \cite{KarafyllisI2018,Huang2020}. For instance, the literature \cite{Xing2017} proposed an adaptive event-triggered control method with a switching threshold strategy to address the control problem of uncertain strict-feedback nonlinear systems. The study in reference \cite{XLT20173} extended the results of reference \cite{Xing2017} to accommodate actuator faults. The work in reference \cite{HuangYX2019} investigated the global event-triggered tracking problem for a class of uncertain nonlinear systems, focusing on reducing data transmission in the sensor-to-controller and controller-to-actuator channels. In reference \cite{KarafyllisI2018}, a novel scheme for triggering parameter estimation was introduced to address the adaptive certainty equivalent control problem of nonlinear systems. The literature \cite{LongJ2022} examined the consensus problem of uncertain high-order nonlinear multi-agent systems with parameter uncertainties and event-triggered communication. An adaptive event-triggered control scheme is proposed for a class of uncertain nonlinear systems where the controller and parameter estimator are event-triggered, ultimately ensuring asymptotic convergence of the system states \cite{Huang2020}.

The aforementioned studies have successfully solved the adaptive event-triggered control problem for nonlinear systems with time-invariant parameters. However, in practical engineering, many parameters may vary with time, such as the aerodynamic stiffness in the wing flutter problem, where aerodynamic heating causes the stiffness and damping of the material to change with time when the aircraft speed changes drastically \cite{LiuCS2017}, and the mass of electromagnetic suspension varying due to passenger movement \cite{OHSY2020}. For such real systems with time-varying parameters, traditional adaptive event-triggered control methods may fail to guarantee the required control performance or even maintain system stability. Therefore, it is essential to study adaptive event-triggered control of systems with time-varying parameters. Recently, the literature \cite{YeHF20232} successfully established the adaptive prescribed-time control scheme for parameter-varying systems. However, such a method is difficult to be compatible with event-triggered control.

Through the above analysis, we recognize that there are three urgent problems and challenges to be solved in studying adaptive event-triggered control for parameter-varying nonlinear systems: i) how to handle time-varying parameters with unknown variation boundaries in the feedback path and output path to achieve zero-error convergence; ii)How to deal with the perturbation term arising from the coupling of unknown input gain and system states, which usually cannot be compensated by designing damping terms because it will produce an algebraic loop; iii) and how to integrate event-triggering mechanisms into the developed adaptive framework, thereby reducing the high sampling frequency and saving communication resources.

To address the above problems and challenges, our work integrates adaptive backstepping design \cite{1995Krstic} with nonlinear damping design \cite{Khalil2002} to estimate time-varying parameters through the application of the improved \textit{congelation of variables} method. Specifically, the design process consists of three main steps: (1) an unknown constant $\ell_\theta$ is introduced to avoid the time-derivation that is produced by the time-varying parameter $\theta(t)$ in each step, and an unknown variable $\delta_\theta$ is introduced to upper-bound the inconsistent between $\theta(t)$ and $\ell_\theta$ so that damping terms can be designed to dominate the uncertainties; (2) three adaptive laws for parameter estimation are constructed, among which two are used for estimating the time-varying parameters in the feedback path, one is used for estimating the time-varying parameters in the input path, and two sets of tuning functions are introduced to avoid over-parameterization; and (3) a soft sign function is utilized to handle the disturbance caused by sampling errors to achieve asymptotic stability and avoid the so-called parameter drift. The main contributions of this paper are summarized as follows:
	\begin{itemize}
		\item 
		A Zeno-free event-triggered asymptotic adaptive control approach is proposed for strict-feedback systems with completely unknown time-varying parameters, which incorporates a sampling error compensation mechanism in the control strategy to achieve global convergence of all closed-loop signals while avoiding parameter drift.
		\item Compared to the assumption in the literature \cite{ WCY2000, WXY2001}, the one imposed in this article is more general, as the restricted condition in these works that requires the a priori knowledge of the radius of $\Theta_0$ ($\Theta_0$ denotes some compact set associated with $\theta(t)$) is removed.
		\item A two-level estimation approach for time-varying parameters $\theta(t)$ is utilized, which effectively eliminates the need for prior knowledge of $\delta_{\theta}$, as covered in the model in \cite{Ye2022,Astolfi2021} as a specific and simple case.
	\end{itemize} 
The remainder of this article is organized as follows. Section \ref{sec-2} presents the problem statement, the system model, and basic preliminaries that used throughout this article. An adaptive event-triggered control strategy is presented in Section \ref{sec-3} followed by the stability analysis in Section \ref{sec-4}. Section \ref{sec-5} demonstrates simulation example to illustrate the effectiveness of the proposed controller. Finally, Section \ref{sec-6} summarizes the article and outline future research directions.

\textit{Notations:} $\mathbb{N}$ represents the field of natural numbers. $\mathbb{R}$ represents the field of reals, $\mathbb{R}_+=\left\{a\in\mathbb{R}:a\textgreater0\right\}$. $\mathbb{R}^n$ denotes the $n$-dimensional Euclidean space, while $\mathbb{R}^{n\times m}$ represents the set of $n \times m$ real matrices. The statement $\Gamma\succ0$ indicates that the symmetric matrix $\Gamma$ with suitable dimensions is positive definite. Given a vector $w$, $w^{\top}$ and $|w|$ denote the transpose  and the Euclidean norm of $w$, respectively. Given an $n\times m$ matrix $W$, $W^{\top}$, $|W|$ denotes the transpose  and the Euclidean norm of $W$, respectively, $W\succ0$ indicates that $W$ is positive definite, $|W|_\mathsf{F}=\sqrt{\sum_{i=1}^{n}\sum_{j=1}^{m}W^2_{ij}}$ denotes the Frobenius norm of $W$. 
	
\section{System description and preliminaries} \label{sec-2}
	\subsection{Problem formulation}
Consider the following strict-feedback nonlinear systems with time-varying parameters in the feedback path and the input path:
\begin{equation}\label{system1}
	\left\{\begin{array}{ll}  
		\dot{x}_1=x_2+\phi_1^\top({x}_1)\theta(t),\\
	~~~~\vdots\\
		\dot{x}_i =x_{i+1}+\phi_i^\top(\underline{x}_i)\theta(t),\quad i=2, \cdots, n-1,\\
		~~~~\vdots\\
		\dot{x}_n =b(t)u(t)+\phi_n^\top(\underline{x}_n)\theta(t),\\
	\end{array}\right.
\end{equation} 
where $\underline{x}_i=[x_1,\cdots,x_i]^{\top}\in\mathbb{R}^i, ~i=2, \cdots, n$ are the system states,  $u(t)\in\mathbb{R}$ is the control input, $\theta(t)\in\mathbb{R}^q$ is a vector composed of uncertain time-varying parameter, and $b(t)\in\mathbb{R}$ is the time-varying control coefficient, the regressors $\phi_i(\underline{x}_i): \mathbb{R}^i\rightarrow\mathbb{R}^q$ are smooth mappings and satisfy $\phi_i(0)=0$.
This article aims to design a Zeno-free adaptive event-triggered controller for system (\ref{system1}) such that the closed-loop system is asymptotically stable.

To achieve such an objective, the following assumptions are imposed.
\begin{assumption}\label{assumption1}
	The parameter $\theta(t)$ is piecewise continuous and $\theta(t)\in\Theta_0$, for all $t\textgreater0$, where $\Theta_0$ is an unknown compact set with an unknown radius. The ``radius'' of $\Theta_0$, denoted by $\delta_{\theta}$, is also unknown.
\end{assumption}
\begin{assumption}\label{assumption2}
	 We assume that the control direction is known and does not change, and there exist an unknown constant $\ell_b$ and a known constant $\bar{b}$, such that $0\textless|\ell_b|\leq|b(t)|\leq\bar{b}$.
\end{assumption}

\begin{remark}
	Assumption \ref{assumption1} is naturally satisfied for any practical systems, which makes the controller suitable for situations
	where the system parameters, even the boundary, are not available. The condition imposed on the parameters in the feedback path, as stated in Assumption \ref{assumption1}, broadens the scope of the model beyond that considered in references \cite{LongJ2022,XLT20172,1995Krstic,DingZ2000}, since the latter assumes that $\theta(t)$ be time-invariant. Compared with methods in the literature that require the a priori knowledge of the compact set \cite{WCY2000,WXY2001} or the radius of the compact set \cite{Ye2022,Astolfi2021}, the algorithm introduced in this paper does not demand any a priori information about $\theta(t)$, making it more general. It is also makes the model more general than the one considered in references \cite{YeHF2022} as the latter requires that $\delta_{{\theta}}$ be known. Assumption \ref{assumption2} is slightly more restrictive than references \cite{Ye2022,Astolfi2021,YeHF2022}, as it requires that the time-varying parameter $b(t)$ has a known upper bound that will be used to handle the sampled errors caused by event-triggered mechanism.  
\end{remark}
\begin{lemma}[See the work of Ye and Song. \cite{YeHF20222}]\label{lemma1}
	Given any smooth function $\sigma(t):[0,\infty)\rightarrow[0, \infty)$, the following inequality holds
	\begin{equation}
		|s|-\frac{s^2}{\sqrt{s^2+\sigma^2}}\leq\sigma,\quad \forall s\in\mathbb{R}.
	\end{equation}
\end{lemma}
\subsection{Event-triggering mechanism} 
To reduce data transmission from the controller to the actuator, the event-triggering mechanism is designed as
\begin{equation}\label{tt3}
	u(t)=u_e(t_k), \quad \forall t\in[t_k, t_{k+1}),
\end{equation}
\begin{equation}\label{tt4}
	t_{k+1}=\inf\{t>t_k:|u(t)-u_e(t)|\geq \gamma_u\},\quad t_1=0,
\end{equation}
where, for $k\in\mathbb{N}$, $t_k$ is the event-triggering instant, $\gamma_u\in\mathbb{R}_+$ is the threshold of the triggering event, and $u_e(t)$ is the designed input. Note that, at $t_{k+1}$, the control signal $u(t)$ is updated to $u_e(t_{k+1})$, while in the time interval $t\in[t_k, t_{k+1})$, $u(t)$ remains constant, i.e., $u(t)=u_e(t_k)$. In other words, when the magnitude of the measurement error $|u(t)-u_e(t)|$ exceeds the prescribed threshold $\gamma_u$, the time will be marked as $t_{k+1}$ and the sampled value $u_e(t_{k+1})$ will be transmitted to the controller, which has the potential of reducing the frequency of the transmission of the signal and the execution of the actuator. 
	
\section{EVENT-TRIGGERED CONTROLLER DESIGN}\label{sec-3}
\begin{figure}[!]
	\centering 
	\includegraphics[width=3.5in]{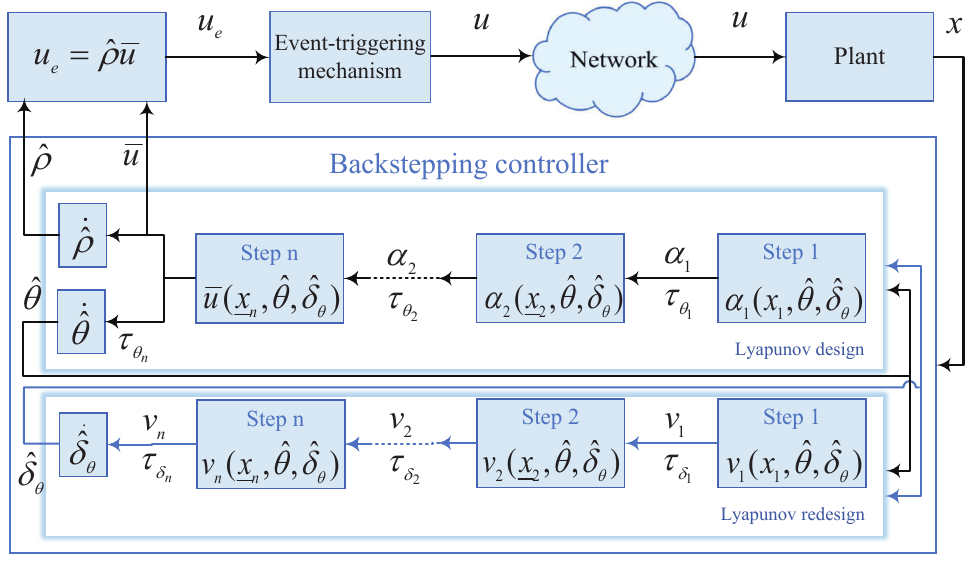}
	\caption{The block diagram of the closed-loop system. The virtual control inputs $\alpha_1,\dots,\alpha_{n}$ and the tuning functions $\tau_{\theta_1},\dots,\tau_{\theta_n}$ are designed in Section \ref{sec-3.2}, the nonlinear damping terms $v_1,\dots,v_n$ and the tuning functions $\tau_{\delta_1},\dots, \tau_{\delta_n}$ are designed in Section \ref{sec-tuning}, the adaptive parameters ${\hat{\theta}}$, ${\hat{\rho}}$, and ${\hat{\delta}}_{\theta}$ are updated by (\ref{tt2}), (\ref{tt12}), and (\ref{tt21}), respectively, the virtual control input $\bar{u}$ is constructed in (\ref{otrr11}), the designed input $u_e$ are formulated in (\ref{otrr113}), and the output of the actuator $u$ is generated through the event-triggered mechanism (\ref{tt3}). }	\label{fig-7}
\end{figure}

This section is devoted to establishing an adaptive event-triggered asymptotic control scheme for system (\ref{system1}) under the architecture depicted by Fig. \ref{fig-7}. Specifically, it is organized into three parts, namely: 1) adaptive estimation of the ``average'' of the time-varying parameter through the backstepping-based Lyapunov design; 2) nonlinear damping adaptive design for re-estimation of the discrepancy between time-varying parameters and their ``average''; and 3) negative feedback design-based generation of event-triggered controllers.

To begin with, we define the new regressor vectors as
\begin{equation}\label{omega_defin}
	\omega_i(\underline{x}_i,\hat{\theta},\hat{\delta}_{\theta})=\phi_i-\sum_{j=1}^{i-1}\frac{\partial \alpha_{i-1}}{\partial x_j}\phi_j, \quad i=1,2,\cdots,n,
\end{equation}
where $\alpha_{i}~(i=1,2,\cdots,n)$ are virtual control laws to be designed in (\ref{otr3}), (\ref{otrr2}), and (\ref{ttt4}).
 
We employ the following equations to perform a decomposition of time-varying parameters $\theta(t)$ and $b(t)$, taking $\Delta_\theta=\theta(t)-\ell_\theta$, $ \Delta_b=b(t)-\ell_b$ are unknown time-varying terms, and $u_e=\hat{\rho}\bar{u}$,
\begin{equation}
	\theta(t)=\hat{\theta}+(\ell_\theta-\hat{\theta})+\Delta_\theta,
\end{equation}
\begin{equation}\label{eqution-7}
	b(t)u=b(t)(u-u_e)+\bar{u}-\ell_b\left(\frac{1}{\ell_b}-\hat{\rho}\right)\bar{u}+\hat\rho\bar{u}\Delta_b,
\end{equation}
where $\ell_\theta$ can be regarded as the ``average'' of $\theta(t)$ (see Reference \cite{Astolfi2021}), which is not necessarily known. Additionally, $\ell_b$ is defined in Assumption \ref{assumption2}, and $\hat{\rho}$ is an ``estimate'' of $1/\ell_b$.
 
\subsection{Lyapunov design by congealed variables and tuning functions}\label{sec-3.2}
In this subsection, we present a control scheme for the strict-feedback system based on congealed variables and tuning functions. In particular, define the coordinate transformations $z_i~(i=1,2,\cdots,n)$ as
\begin{equation}\label{z_1}
	z_1=x_1,
\end{equation}
\begin{equation}\label{z_i}
	z_i=x_i-\alpha_{i-1}(\underline{x}_{i-1},\hat{\theta},\hat{\delta}_{\theta}),\quad i=2,\cdots,n,
\end{equation}
where $\alpha_{i}~(i=1,2,\cdots,n)$ refers to virtual controllers. The details design procedure is given in the following steps.

$\mathbf{{\it Step~~ 1 }}$ : To stabilize subsystem $\dot{x}_1=x_2+\phi^{\top}_1(x_1)\theta(t)$, we choose the Lyapunov function candidate as
\begin{equation}\label{otr1}
	\begin{aligned}
		V_1=\frac{1}{2}z_1^2+\frac{1}{2}(\ell_\theta-\hat{\theta})^\top\Gamma^{-1}(\ell_\theta-\hat{\theta}),
	\end{aligned}
\end{equation}
where $\Gamma\succ0$ is a positive definite matrix. Taking the time derivative of ${V}_1$ along the trajectories of (\ref{z_1}), we have
\begin{equation}\label{otr2}
	\begin{aligned}
		\dot{V_1}&=z_1(\alpha_1+z_2+\omega^{\top}_1\hat{\theta}+\omega^{\top}_1\Delta_\theta)+(\ell_\theta-\hat{\theta})^{\top}\Gamma^{-1}(\Gamma\omega_1z_1-\dot{\hat{\theta}}).
	\end{aligned}
\end{equation}
Next, we design the virtual control law $\alpha_1(x_1,\hat{\theta},\hat{\delta}_{\theta})$ and the tuning function $\tau_{\theta_1}$ as
\begin{equation}\label{otr3}
	\begin{aligned}
		\alpha_1=-k_1z_1+v_1(x_1,\hat{\theta},\hat{\delta}_{\theta})-\omega^{\top}_1\hat{\theta},
	\end{aligned}
\end{equation}
\begin{equation}\label{tt10}
	\begin{aligned}
		\tau_{\theta_1}=\omega_1z_1,
	\end{aligned}
\end{equation}
where $k_1\textgreater0$, and $\omega_1=\phi_1$ is defined in (\ref{omega_defin}), and $v_1(x_1,\hat{\theta},\hat{\delta}_{\theta})$ is a nonlinear damping term to be designed in (\ref{tt14}).
Substituting (\ref{otr3}) and (\ref{tt10}) into (\ref{otr2}), yields
\begin{equation}\label{otrr1}
	\begin{aligned}
		\dot{V_1}=&-k_1z^2_1+z_1(v_1+\omega^{\top}_1\Delta_\theta+z_2)+(\ell_\theta-\hat{\theta})^\top\Gamma^{-1}(\Gamma\tau_{\theta_1}-\dot{\hat{\theta}}).
	\end{aligned}
\end{equation}

$\mathbf{{\it Step~~2 }}$ : Consider subsystem $\dot{x}_2=x_3+\phi^{\top}_2(x_1,x_2)\theta(t)$ and recalling that $z_2=x_2-\alpha_{1}(x_1,\hat{\theta},\hat{\delta}_{\theta})$, the dynamics of $z_2$ can be shown as:
\begin{equation}
	\dot{z}_2=\alpha_2+z_{3}+\omega^{\top}_2\hat{\theta}+\omega^{\top}_2\Delta_\theta+\omega^{\top}_2(\ell_\theta-\hat{\theta})-\frac{\partial \alpha_1}{\partial x_1}x_2-\frac{\partial {\alpha}_{1}}{\partial \hat{\theta}}\dot{\hat{\theta}}-\frac{\partial {\alpha}_{1}}{\partial \hat{\delta}_\theta}\dot{\hat{\delta}}_\theta.
\end{equation}
Choose the Lyapunov function candidate as $V_2=V_1+\frac{1}{2}z_2^2$. Its derivative is given by:
\begin{equation}\label{otr5}
	\begin{aligned}
		\dot{V_2}=&\dot{V}_1+z_2\left(\alpha_2+z_3+\omega^{\top}_2\hat{\theta}+\omega^{\top}_2\Delta_\theta-\frac{\partial \alpha_1}{\partial x_1}x_2\right)+z_2\omega^{\top}_2(\ell_\theta-\hat{\theta})-z_2\left(\frac{\partial \alpha_1}{\partial \hat{\theta}}\dot{\hat{\theta}}+\frac{\partial \alpha_1}{\partial \hat{\delta}_{\theta}}\dot{\hat{\delta}}_\theta\right).
	\end{aligned}
\end{equation}
Then, we design the virtual control law $\alpha_2(\underline{x}_{2},\hat{\theta},\hat{\delta}_{\theta})$ and the tuning function $\tau_{\theta_2}$ as
\begin{equation}\label{otrr2}
	\begin{aligned}
		\alpha_2=-k_2z_2+v_2(\underline{z}_2,\hat{\delta}_{\theta})-\omega^{\top}_2\hat{\theta}+\frac{\partial \alpha_1}{\partial x_1}x_2+\frac{\partial \alpha_1}{\partial \hat{\theta}}\Gamma\tau_{\theta_2}-z_1,
	\end{aligned}
\end{equation}
\begin{equation}\label{ttt1}
	\begin{aligned}
		\tau_{\theta_2}=\tau_{\theta_1}+\omega_2z_2=\sum_{j=1}^{2}\omega_jz_j,
	\end{aligned}
\end{equation}
where $k_2\textgreater0$, and $\omega_2=\phi_2-\frac{\partial \alpha_{1}}{\partial x_1}\phi_1$ is defined in (\ref{omega_defin}), and $v_2(\underline{x}_2,\hat{\theta},\hat{\delta}_{\theta})$ is a nonlinear damping term to be designed in (\ref{otrr8}). By substituting (\ref{otrr2}) and (\ref{ttt1}) into (\ref{otr5}), yields
\begin{equation}
	\begin{aligned}
		\dot{V_2}=&-\sum_{j=1}^{2}k_jz^2_j+z_2z_3+\sum_{j=1}^{2}(z_jv_j+z_j\omega^{\top}_j\Delta_\theta)+\left(\frac{\partial \alpha_1}{\partial \hat{\theta}}z_2+(\ell_\theta-\hat{\theta})^{\top}\Gamma^{-1}\right)(\Gamma\tau_{\theta_2}-\dot{\hat{\theta}})-z_2\frac{\partial \alpha_1}{\partial \hat{\delta}_{\theta}}\dot{\hat{\delta}}_{\theta}.\\
	\end{aligned}
\end{equation}
$\mathbf{{\it Step~~ i~(i=3,\cdots,n-1)}}$: Based on the results of the previous steps and with the consideration of the function $V_i=V_{i-1}+\frac{1}{2}z^2_i$, the virtual control inputs and adaptive laws can be recursively obtained by following the standard backstepping procedure. Recalling (\ref{z_i}), the dynamics of $z_i$ can be shown as:
\begin{equation}\label{dotz_i}
	\dot{z}_i=z_{i+1}+\alpha_i+\omega^{\top}_i\hat{\theta}+\omega^{\top}_i\Delta_\theta+\omega^{\top}_i(\ell_\theta-\hat{\theta})-\sum_{j=1}^{i-1}\frac{\partial \alpha_{i-1}}{\partial x_j}x_{j+1}-\frac{\partial {\alpha}_{i-1}}{\partial \hat{\theta}}\dot{\hat{\theta}}-\frac{\partial {\alpha}_{i-1}}{\partial \hat{\delta}_\theta}\dot{\hat{\delta}}_\theta.
\end{equation}
Taking the time derivative of ${V}_i$ along the trajectories of (\ref{dotz_i}), we obtain
\begin{equation}\label{tt11}
	\begin{aligned}
		\dot{V}_i=&\dot{V}_{i-1}+z_i\left(\alpha_i+z_{i+1}+\omega^{\top}_i\hat{\theta}+\omega^{\top}_i\Delta_\theta-\sum_{j=1}^{i-1}\frac{\partial \alpha_{i-1}}{\partial x_j}x_{j+1}\right)+z_i\omega^{\top}_i(\ell_\theta-\hat{\theta})-z_i\left(\frac{\partial \alpha_{i-1}}{\partial \hat{\theta}}\dot{\hat{\theta}}+\frac{\partial \alpha_{i-1}}{\partial \hat{\delta}_{\theta}}\dot{\hat{\delta}}_{\theta}\right).
	\end{aligned}
\end{equation}
Next, we design the virtual control law $\alpha_i(\underline{x}_{i},\hat{\theta},\hat{\delta}_{\theta})$ and the tuning function $\tau_{\theta_i}$ as
\begin{equation}\label{ttt4}
	\begin{aligned}
		\alpha_i=&-k_iz_i+v_i(\underline{x}_i,\hat{\theta},\hat{\delta}_{\theta})-\omega^{\top}_i\hat{\theta}+\sum_{j=1}^{i-1}\frac{\partial \alpha_{i-1}}{\partial x_j}x_{j+1}+\frac{\partial \alpha_{i-1}}{\partial \hat{\theta}}\Gamma\tau_{\theta_i}+\sum_{j=2}^{i-1}\frac{\partial \alpha_{j-1}}{\partial \hat{\theta}}\Gamma\omega_iz_j-z_{i-1},
	\end{aligned}
\end{equation}
\begin{equation}\label{ttt2}
	\begin{aligned}
		\tau_{\theta_i}=\tau_{\theta_{i-1}}+\omega_iz_i=\sum_{j=1}^{i}\omega_jz_j,
	\end{aligned}
\end{equation}
where $k_i\textgreater0$, and $\omega_i$ is defined in (\ref{omega_defin}), and $v_i(\underline{x}_i,\hat{\theta},\hat{\delta}_{\theta})$ is a nonlinear damping term to be designed in (\ref{tt17}).
Then, (\ref{ttt4}), (\ref{ttt2}) together with (\ref{tt11}), implies that
\begin{equation}\label{dotV_i}
	\begin{aligned}
		\dot{V}_i=&-\sum_{j=1}^{i}k_jz^2_j+z_iz_{i+1}+\sum_{j=1}^{i}(z_jv_j+z_j\omega^{\top}_j\Delta_\theta)+\left(\sum_{j=1}^{i-1}\frac{\partial \alpha_j}{\partial \hat{\theta}}z_{j+1}+(\ell_\theta-\hat{\theta})^{\top}\Gamma^{-1}\right)(\Gamma\tau_{\theta_i}-\dot{\hat{\theta}})-z_i\frac{\partial \alpha_{i-1}}{\partial \hat{\delta}_{\theta}}\dot{\hat{\delta}}_{\theta}.
	\end{aligned}
\end{equation}
$\mathbf{{\it Step~~n}}$ : Consider subsystem $\dot{x}_n =b(t)u(t)+\phi_n^\top(\underline{x}_n)\theta(t)$ and recalling that $z_n=x_n-\alpha_{n-1}(\underline{x}_{n},\hat{\theta},\hat{\delta}_{\theta})$, the dynamics of $z_n$ is given by:
\begin{equation}\label{ottr5}
	\begin{aligned}
		\dot{z}_n&
		=b(t)(u-u_e)+\bar{u}-\ell_b\left(\frac{1}{\ell_b}-\hat{\rho}\right)\bar{u}+\hat\rho\bar{u}\Delta_b+\omega^{\top}_n\hat{\theta}+\omega^{\top}_n\Delta_\theta+\omega^{\top}_n(\ell_\theta-\hat{\theta})\\
		&~~~~-\sum_{j=1}^{n-1}\frac{\partial \alpha_{n-1}}{\partial x_j}x_{j+1}-\frac{\partial {\alpha}_{n-1}}{\partial \hat{\theta}}\dot{\hat{\theta}}-\frac{\partial {\alpha}_{n-1}}{\partial \hat{\delta}_\theta}\dot{\hat{\delta}}_\theta.
	\end{aligned}
\end{equation}
In this case, a candidate Lyapunov function $V_n$ can be chosen as
\begin{equation}\label{otrr3}
	V_n=V_{n-1}+\frac{1}{2}z^2_{n}+V_\rho,
\end{equation}
where $V_\rho=\frac{|\ell_b|}{2\gamma_\rho}\left(\frac{1}{\ell_b}-\hat{\rho}\right)^2$, with $\gamma_\rho\textgreater0$ being a constant. Taking the time derivative of $V_n$ along the trajectories of (\ref{ottr5}), we obtain
\begin{equation}\label{tt13}
	\begin{aligned}
		\dot{V_n}=&\dot{V}_{n-1}+z_n\left(\bar{u}-\sum_{j=1}^{n-1}\frac{\partial \alpha_{n-1}}{\partial x_j}x_{j+1}+\omega^{\top}_n\hat{\theta}+\omega^{\top}_n\Delta_\theta\right)+z_n\omega^{\top}_n(\ell_\theta-\hat{\theta})+z_n\bar{u}\hat{\rho}\Delta_b\\
		&-z_n\left(\frac{\partial \alpha_{n-1}}{\partial \hat{\theta}}\dot{\hat{\theta}}+\frac{\partial\alpha_{n-1}}{\partial \hat\delta_{{\theta}}}\dot{\hat{\delta}}_{\theta}\right)-\frac{|\ell_b|}{\gamma_\rho}\left(\frac{1}{\ell_b}-\hat{\rho}\right)\left(\gamma_\rho\sign{\ell_b}z_n\bar{u}+\dot{\hat{\rho}}\right)+b(u-u_e)z_n.
	\end{aligned}
\end{equation}
Note that the last term $b(u-u_e)z_n$ represents the disturbance generated by the sampling error $u-u_e$ due to event-triggering, which will be processed in (\ref{bt_1}).
Next, design the control input $\bar{u}(\underline{x}_{n},\hat{\theta},\hat{\delta}_{\theta})$ and the tuning function $\tau_{\theta_n}$ as
\begin{equation}\label{baru-defin}
	\bar{u}=-k_nz_n+v_n(\underline{x}_n,\hat{\theta},\hat{\delta}_{\theta})+\kappa z_n,
\end{equation}
\begin{equation}\label{tau_n}
	\tau_{\theta_n}=\tau_{\theta_{n-1}}+\omega_nz_n=\sum_{j=1}^{n}\omega_jz_j,
\end{equation}
where $k_n\textgreater0$ holds, and $\omega_n=\phi_n-\sum_{j=1}^{n-1}\frac{\partial \alpha_{n-1}}{\partial x_j}\phi_j$ is defined in (\ref{omega_defin}). Additionally, $v_n(\underline{x}_n,\hat{\theta},\hat{\delta}_{\theta})$ is a nonlinear damping term to be designed in (\ref{tt19}), and $\kappa>0$ is a positive state-dependent function to be designed in (\ref{ttt26}).

With the tuning functions (\ref{tt10}), (\ref{ttt1}), (\ref{ttt2}) and (\ref{tau_n}), the parameter update law can be designed as
\begin{equation}\label{tt2}
	\begin{aligned}
		&\dot{\hat{\theta}}=\Gamma\tau_{\theta_n}=\Gamma\sum_{i=1}^{n}\omega_iz_i,\\
	\end{aligned}
\end{equation}
\begin{equation}\label{tt12}
	\dot{\hat{\rho}}=-\gamma_\rho\sign{\ell_b}\bar{u}z_n.
\end{equation}
Substituting (\ref{dotV_i}), and (\ref{baru-defin})-(\ref{tt12}) into (\ref{tt13}), we have
\begin{equation}\label{ttv_n}
	\begin{aligned}
		\dot{V_n}=&-\sum_{j=1}^{n}k_jz^2_j+z_n\left(z_{n-1}+\omega^{\top}_n\hat{\theta}+\kappa z_n-\sum_{j=1}^{n-1}\frac{\partial \alpha_{n-1}}{\partial x_j}x_{j+1}-\sum_{j=2}^{n-1}\frac{\partial \alpha_{j-1}}{\partial \hat{\theta}}\Gamma z_j\omega_n-\frac{\partial \alpha_{n-1}}{\partial \hat{\theta}}\Gamma\tau_{\theta_n}\right)\\
		&-\sum_{j=2}^{n}z_j\frac{\partial\alpha_{j-1}}{\partial \hat{\delta}_{\theta}}\dot{\hat{\delta}}_{\theta}+z_n\bar{u}\hat{\rho}\Delta_b+\sum_{j=1}^{n}\left(z_jv_j+z_j\omega^{\top}_j\Delta_\theta\right)+b(u-u_e)z_n.
	\end{aligned}
\end{equation}

\subsection{Lyapunov redesign by nonlinear damping and tuning functions}\label{sec-tuning}
In this subsection, we proposed an adaptive scheme to handle $\Delta_{\theta}$ through a Lyapunov redesign. Initially, by utilizing Hadamard's lemma (see Reference \cite{Nestruev2006}), we can derive 
\begin{equation}\label{omega_w}
	\omega_i=\textit{W}^\top_i \underline{z}_i,
\end{equation}
 where $\underline{z}_i=[z_1, z_2, \cdots,z_i]^{\top}$, and $\textit{W}_i\in\mathbb{R}^{i\times q}$ represent the smooth mappings. Furthermore, by applying Young’s inequality and Assumption \ref{assumption1}, the following inequality can be obtained:
\begin{equation}\label{ttt22}
	\begin{aligned}
		z_i\omega^{\top}_i\Delta_\theta\leqslant\frac{\delta_{\theta}}{2}(|W_i|_{\mathsf{F}}^2+n+1-i)z^2_i+
		\frac{\delta_{\theta}}{2}\underline{z}^{\top}_{i-1}\underline{z}_{i-1},~~ i=1,\cdots,n,
	\end{aligned}
\end{equation}
where $\underline{z}^{\top}_{i-1}\underline{z}_{i-1}=z^2_1+z^2_2+\cdots+z^2_{i-1}$ is used. The following steps outline the specific process for Lyapunov redesign.

$\mathbf{{\it Step~~1}}$: We choose the candidate Lyapunov function as 
\begin{equation}
	V_\delta=\frac{1}{2\gamma_\delta}(\delta_{\theta}-\hat{\delta}_{\theta})^2,
\end{equation} 
where $\gamma_\delta\textgreater0$. The nonlinear damping term $v_1(x_1,\hat{\theta},\hat{\delta}_{\theta})$ and the tuning function $\tau_{\delta_1}$ can be designed as follows:
\begin{equation}\label{tt14}
	\begin{aligned}
		v_1&=-\frac{\hat{\delta}_{\theta}}{2}\left(|W_1|^{2}_{\mathsf{F}}+n\right)z_1-\frac{1}{2\epsilon_\Omega}z_1,
	\end{aligned}
\end{equation}
\begin{equation}\label{tt15}
	\tau_{\delta_1}=\frac{1}{2}\left(|W_1|^{2}_{\mathsf{F}}+n\right)z^2_1,
\end{equation}
where $\epsilon_\Omega\textgreater0$, and $W_1\in\mathbb{R}^{1\times q}$ is a smooth mapping that satisfies $\omega_1=W^{\top}_1 z_1$. Using (\ref{ttt22}), (\ref{tt14}) and (\ref{tt15}), we have
\begin{equation}\label{ttt23}
	\begin{aligned}
		&z_1v_1+z_1\omega^{\top}_1\Delta_\theta+\dot{V}_\delta\leq\frac{\tilde{\delta}_{\theta}}{\gamma_\delta}\left(\gamma_\delta\tau_{\delta_1}-\dot{\hat{\delta}}_{\theta}\right)-\frac{1}{2\epsilon_\Omega}z^2_1,
	\end{aligned}
\end{equation}
where $\tilde{\delta}_{\theta}=\delta_{\theta}-\hat{\delta}_{\theta}$. Combining (\ref{otr3}) and (\ref{tt14}), we can express the virtual control $\alpha_1$ as
\begin{equation}\label{alpha_1}
	\alpha_1=-k_1z_1-\omega^{\top}_1\hat{\theta}-\frac{\hat{\delta}_{\theta}}{2}\left(|W_1|^{2}_{\mathsf{F}}+n\right)z_1-\frac{1}{2\epsilon_\Omega}z_1.
\end{equation}
$\mathbf{{\it Step~~2}}$: The nonlinear damping term $v_2(\underline{x}_2,\hat{\theta},\hat{\delta}_{\theta})$ and the tuning function $\tau_{\delta_2}$ can be designed as
\begin{equation}\label{otrr8}
	\begin{aligned}
		&v_2=-\frac{\hat{\delta}_{\theta}}{2}\left(|W_2|^2_{\mathsf{F}}+n-1\right)z_2-\frac{1}{2\epsilon_\Omega}z_2+\gamma_\delta\frac{\partial \alpha_1}{\partial \hat{\delta}_{\theta}}\tau_{\delta_2},\\
	\end{aligned}
\end{equation}
\begin{equation}\label{tt16}
	\tau_{\delta_2}=\tau_{\delta_1}+\frac{1}{2}\left(|W_2|^2_{\mathsf{F}}+n-1\right)z^2_2,
\end{equation}
where $W_2\in\mathbb{R}^{2\times q}$ is a smooth mapping that satisfies $\omega_2=W^{\top}_2 z_2$, and we can utilize (\ref{ttt22}), (\ref{ttt23}), (\ref{otrr8}), and (\ref{tt16}) to obtain:
\begin{equation}\label{ttdelt_n}
	\sum_{j=1}^{2}(z_jv_j+z_j\omega^{\top}_j\Delta_\theta)+\dot{V}_\delta\leq\frac{\tilde{\delta}_{\theta}}{\gamma_\delta}(\gamma_\delta\tau_{\delta_2}-\dot{\hat{\delta}}_{\theta})+\gamma_\delta\frac{\partial \alpha_1}{\partial \hat{\delta}_{\theta}}z_2\tau_{\delta_2}-\frac{1}{2\epsilon_\Omega}\sum_{j=1}^{2}z^2_i.
\end{equation}
By integrating \ref{otrr2}) and (\ref{otrr8}), the virtual control $\alpha_2$ can be stated as
\begin{equation}\label{otrr9}
	\begin{aligned}
		\alpha_2=&-k_2z_2-\omega^{\top}_2\hat{\theta}-\frac{\hat{\delta}_{\theta}}{2}\left(|W_2|^2_\mathsf{F}+n-1\right)z_2-\frac{1}{2\epsilon_\Omega}z_2+\gamma_\delta\frac{\partial \alpha_1}{\partial \hat{\delta}_{\theta}}\tau_{\delta_2}+\frac{\partial \alpha_1}{\partial x_1}x_2+\frac{\partial \alpha_1}{\partial \hat{\theta}}\Gamma\tau_{\theta_2}-z_1.
	\end{aligned}
\end{equation}

$\mathbf{{\it Step~~i~~(i=3,\cdots,n-1)}}$ : Utilizing the design techniques discussed above, we formulated the nonlinear damping term $v_i(\underline{x}_i,\hat{\theta},\hat{\delta}_{\theta})$ and the tuning function $\tau_{\delta_i}$ as:
\begin{equation}\label{tt17}
	\begin{aligned}
		v_i=&-\frac{\hat{\delta}_{\theta}}{2}\left(|W_i|^2_\mathsf{F}+n+1-i\right)z_i-\frac{1}{2\epsilon_\Omega}z_i+\gamma_\delta\frac{\partial \alpha_{i-1}}{\partial \hat{\delta}_{\theta}}\tau_{\delta_i}+\gamma_\delta\left[\frac{1}{2}\left(|W_i|^2_\mathsf{F}+n+1-i+\frac{1}{\epsilon_\Omega}\right)z_i\right]\sum_{j=2}^{i-1}\frac{\partial \alpha_{j-1}}{\partial \hat{\delta}_{\theta}}z_j,
	\end{aligned}
\end{equation}
\begin{equation}\label{tt18}
	\begin{aligned}
		\tau_{\delta_i}=\tau_{\delta_{i-1}}+\frac{1}{2}\left(|W_i|^2_\mathsf{F}+n+1-i\right)z^2_i,
	\end{aligned}
\end{equation}
where $W_i\in\mathbb{R}^{i\times q}$ is a smooth mapping that satisfies (\ref{omega_w}). By utilizing (\ref{ttt22}), (\ref{ttdelt_n}), (\ref{tt17}), and (\ref{tt18}), we obtain
\begin{equation}\label{ttt222}
	\begin{aligned}
		\sum_{j=1}^{i}(z_jv_j+z_j\omega^{\top}_j\Delta_\theta)+\dot{V}_\delta&\leq\sum_{j=3}^{i}\left[z_jv_j+\frac{\hat{\delta}_{\theta}}{2}\left(|W_j|^2_\mathsf{F}+n+1-j\right)z^2_j\right]+\frac{\tilde{\delta}_{\theta}}{\gamma_\delta}(\gamma_\delta\tau_{\delta_2}-\dot{\hat{\delta}}_{\theta})+\gamma_\delta\frac{\partial \alpha_1}{\partial \hat{\delta}_{\theta}}z_2\tau_{\delta_2}\\
		&=\frac{\tilde{\delta}_{\theta}}{\gamma_\delta}(\gamma_\delta\tau_{\delta_i}-\dot{\hat{\delta}}_{\theta})+\sum_{j=1}^{i}\gamma_\delta\left[\frac{1}{2}\left(|W_j|^2_\mathsf{F}+n+1-j\right)z^2_j\sum_{p=2}^{j-1}\frac{\partial \alpha_{p-1}}{\partial \hat{\delta}_{\theta}}z_p\right]\\
		&~~~~~-\frac{1}{2\epsilon_\Omega}\sum_{j=1}^{i}z^2_j+\sum_{j=2}^{i}\gamma_\delta\frac{\partial \alpha_{j-1}}{\partial \hat{\delta}_{\theta}}z_j\tau_{\delta_j}.
	\end{aligned}
\end{equation}
Then, combining (\ref{ttt4}) and (\ref{tt17}), we express the virtual control $\alpha_i$ as
\begin{equation}\label{otrr14}
	\begin{aligned}
		\alpha_i=&-k_iz_i-\omega^{\top}_i\hat{\theta}-\frac{\hat{\delta}_{\theta}}{2}\left(|W_i|^2_\mathsf{F}+n+1-i\right)z_i-\frac{1}{2\epsilon_\Omega}z_i+\gamma_\delta\frac{\partial \alpha_{i-1}}{\partial \hat{\delta}_{\theta}}\tau_{\delta_i}+\sum_{j=1}^{i-1}\frac{\partial \alpha_{i-1}}{\partial x_j}x_{j+1}+\frac{\partial \alpha_{i-1}}{\partial \hat{\theta}}\Gamma\tau_{\theta_i}\\
		&+\gamma_\delta\left[\frac{1}{2}\left(|W_i|^2_\mathsf{F}+n+1-i+\frac{1}{\epsilon_\Omega}\right)z_i\right]\sum_{j=2}^{i-1}\frac{\partial \alpha_{j-1}}{\partial \hat{\delta}_{\theta}}z_j+\sum_{j=2}^{i-1}\frac{\partial \alpha_{j-1}}{\partial \hat{\theta}}\Gamma\omega_iz_j-z_{i-1}.
	\end{aligned}
\end{equation}
$\mathbf{{\it Step~~n}}$: In this step, we design the nonlinear damping term  $v_n(\underline{x}_n,\hat{\theta},\hat{\delta}_{\theta})$ and the tuning function $\tau_{\delta_n}$ as
\begin{equation}\label{tt19}
	\begin{aligned}
		&v_n=-\frac{\hat{\delta}_{\theta}}{2}\left(|W_n|^2_\mathsf{F}+1\right)z_n-\frac{1}{2\epsilon_\Omega}z_n,
	\end{aligned}
\end{equation}
\begin{equation}\label{tt20}
	\tau_{\delta_n}=\tau_{\delta_{n-1}}+\frac{1}{2}\left(|W_n|^2_\mathsf{F}+1\right)z^2_n,
\end{equation}
where $W_n\in\mathbb{R}^{n\times q}$ is a smooth mapping that satisfies $\omega_n=W^{\top}_n \underline{z_n}$.

Design the update law of $\hat{\delta}_{\theta}$ as
\begin{equation}\label{tt21}
	\dot{\hat{\delta}}_{\theta}=\gamma_\delta\tau_{\delta_n}
	=\gamma_\delta\sum_{i=1}^{n}\frac{1}{2}\left(|W_i|^2_\mathsf{F}+n+1-i\right)z^2_i.
\end{equation}
Next, by employing (\ref{ttt22}), (\ref{ttt222}) and (\ref{tt19})-(\ref{tt21}), we have
\begin{equation}\label{ttw1}
	\begin{aligned}
		&\sum_{j=1}^{n}(z_jv_j+z_j\omega^{\top}_j\Delta_\theta)+\dot{V}_\delta-\sum_{j=2}^{n}z_j\frac{\partial \alpha_{j-1}}{\partial \hat{\delta}_{\theta}}\dot{\hat{\delta}}_{\theta}\leq-\gamma_\delta\frac{\partial \alpha_{n-1}}{\partial \hat{\delta}_{\theta}}z_n\tau_{\delta_n}-\frac{\gamma_\delta}{2}\left(|W_n|^2_\mathsf{F}+1\right)z^2_n\sum_{j=2}^{n-1}\frac{\partial \alpha_{j-1}}{\partial \hat{\delta}_{\theta}}z_j-\frac{1}{2\epsilon_\Omega}\sum_{j=1}^{n}z^2_j.
	\end{aligned}
\end{equation} 
Finally, the Lyapunov function is chosen as   
\begin{equation}
	V=V_z+V_\theta+V_\rho+V_\delta,
\end{equation}
where $V_z=\frac{1}{2}\sum_{j=1}^{n}z^2_j$, and $V_\theta=\frac{1}{2}(\ell_\theta-\hat{\theta})^\top\Gamma^{-1}(\ell_\theta-\hat{\theta})$. The detailed expression for $V$ is presented below:
\begin{equation}\label{ottr1}
	\begin{aligned}
		V=\frac{1}{2}\sum_{j=1}^{n}z^2_j+\frac{1}{2}(\ell_\theta-\hat{\theta})^\top\Gamma^{-1}(\ell_\theta-\hat{\theta})+\frac{|\ell_b|}{2\gamma_\rho}\left(\frac{1}{\ell_b}-\hat{\rho}\right)^2+\frac{1}{2\gamma_\delta}(\delta_{\theta}-\hat{\delta}_{\theta})^2.
	\end{aligned}
\end{equation}
Taking the time derivative of $V$ along the trajectories of (\ref{ottr5}), and recalling (\ref{ttv_n}) and (\ref{ttw1}), we have
\begin{equation}\label{ottr122}
	\begin{aligned}
		\dot{V}&=-\sum_{j=1}^{n}k_jz^2_j+z_n\left(z_{n-1}+\omega^{\top}_n\hat{\theta}-\sum_{j=1}^{n-1}\frac{\partial \alpha_{n-1}}{\partial x_j}x_{j+1}-\sum_{j=2}^{n-1}\frac{\partial \alpha_{j-1}}{\partial \hat{\theta}}\Gamma z_j\omega_n-\frac{\partial \alpha_{n-1}}{\partial \hat{\theta}}\Gamma\tau_{\theta_n}+\kappa z_n\right)+z_n\bar{u}\hat{\rho}\Delta_b\\
		&~~~~~+b(t)(u-u_e)z_n+\sum_{j=1}^{n}(z_jv_j+z_j\omega^{\top}_j\Delta_\theta)+\dot{V}_\delta-\sum_{j=2}^{n}z_j\frac{\partial \alpha_{j-1}}{\partial \hat{\delta}_{\theta}}\dot{\hat{\delta}}_{\theta},\\
		&\leq-\sum_{j=1}^{n}k_jz^2_j+z_n\left(z_{n-1}+\omega^{\top}_n\hat{\theta}-\sum_{j=1}^{n-1}\frac{\partial \alpha_{n-1}}{\partial x_j}x_{j+1}-\Psi_\theta-\Psi_\delta+\kappa z_n\right)+z_n\bar{u}\hat{\rho}\Delta_b+b(t)(u-u_e)z_n-\frac{1}{2\epsilon_\Omega}\sum_{j=1}^{n}z^2_j,
	\end{aligned}
\end{equation}
where
\begin{equation}
	\begin{aligned}
		&\Psi_\theta=\frac{\partial \alpha_{n-1}}{\partial \hat{\theta}}\Gamma\tau_{\theta_n}+\sum_{j=2}^{n-1}\frac{\partial \alpha_{j-1}}{\partial \hat{\theta}}\Gamma  z_j \omega_n,\\
		&\Psi_\delta=\gamma_\delta\frac{\partial \alpha_{n-1}}{\partial \hat{\delta}_{\theta}}\tau_{\delta_n}+\frac{\gamma_\delta}{2}\left(|W_n|^2_\mathsf{F}+1\right)z_n\sum_{j=2}^{n-1}\frac{\partial \alpha_{j-1}}{\partial \hat{\delta}_{\theta}}z_j.
	\end{aligned}
\end{equation}
 The following inequality holds by applying Lemma \ref{lemma1}:
\begin{equation}\label{bt_1}
	\begin{aligned}
		b(t)(u-u_e)z_n\leq |b(t)|\gamma_u |z_n|\leq |\bar{b}|\gamma_u\left(\frac{z^2_n}{\sqrt{z^2_n+\sigma^2}}+\sigma\right),
	\end{aligned}
\end{equation}
where $\sigma=e^{-\xi t}$ with $\xi$ being a positive constant, and $\gamma_u$ denotes the triggering threshold.

Define
\begin{equation}\label{Omega_defin} \Omega=z_{n-1}+\omega^{\top}_n\hat{\theta}-\sum_{j=1}^{n-1}\frac{\partial \alpha_{n-1}}{\partial x_j}x_{j+1}-\Psi_\delta-\Psi_\theta+\frac{|\bar{b}|\gamma_u z_n}{\sqrt{z^2_n+\sigma^2}}.
\end{equation}
Substituting (\ref{bt_1}) and (\ref{Omega_defin}) into (\ref{ottr122}), and (\ref{ottr122}) can be continued as:
\begin{equation}\label{tt1}
	\dot{V}\leq-\sum_{j=1}^{n}k_jz^2_j+z_n
	\bar{u}\hat{\rho}\Delta_b+|\bar{b}|\gamma_u\sigma+z_n\Omega(x,\hat{\theta},\hat{\delta}_{\theta})+\kappa z^2_n-\frac{1}{2\epsilon_\Omega}\sum_{j=1}^{n}z^2_j.
\end{equation}

\subsection{Event-triggered controller generation based on negative feedback design}
In this subsection, we will deal with $\Delta_b$ through negative feedback gain design. Initially, according to Hadamard’s lemma, $\Omega$ can be expressed as $\Omega=\bar{\Omega}^{\top}\underline{z}_n$, with  $\bar{\Omega}\in\mathbb{R}^n$ being a smooth mapping. Then, by applying Young’s inequality with $\epsilon_\Omega>0$, yields
\begin{equation}\label{otrr6}
	\begin{aligned}
		z_n\Omega=z_n\bar{\Omega}^{\top}\underline{z}_n\leq\frac{1}{2}\left(\epsilon_\Omega|\bar{\Omega}|^2+\frac{1}{\epsilon_\Omega}\right)z^2_n+\frac{1}{2\epsilon_\Omega}\underline{z}^{\top}_{n-1}\underline{z}_{n-1},
	\end{aligned}
\end{equation}
where $\underline{z}^{\top}_{n-1}\underline{z}_{n-1}=z^2_1+z^2_2+\cdots+z^2_{n-1}$ .

Design 
\begin{equation}\label{ttt26}
	\begin{aligned}
		&\kappa=-\frac{1}{2}\epsilon_\Omega|\bar{\Omega}|^2.
	\end{aligned}
\end{equation} 
With (\ref{otrr6}) and (\ref{ttt26}), the last three terms of (\ref{tt1}) can be continued as
\begin{equation}\label{ttt25}
	\begin{aligned}
		z_n\Omega+\kappa z^2_n-\frac{1}{2\epsilon_\Omega}\sum_{j=1}^{n}z^2_j=z_n\Omega-\frac{1}{2}\epsilon_\Omega|\bar{\Omega}|^2 z^2_n-\frac{1}{2\epsilon_\Omega}\sum_{j=1}^{n}z^2_j\leq0.
	\end{aligned}
\end{equation}
Inserting (\ref{ttt25}) into (\ref{tt1}), we can obtain
\begin{equation}\label{ttt27}
	\dot{V}\leq-\sum_{j=1}^{n}k_jz^2_j+z_n
	\bar{u}\hat{\rho}\Delta_b+|\bar{b}|\gamma_u\sigma.
\end{equation}

It is worth noting that the perturbation term, $z_n\bar{u}\hat{\rho}\Delta_b$, is explicitly dependent on $\bar{u}$. Merely adding damping terms to $\bar{u}$ to control this term would result in undesired alterations to the perturbation term itself. Rather, designing $\bar{u}$ and selecting $\hat{\rho}(0)$ is vital in making sure that the perturbation term becomes non-positive.
According to (\ref{baru-defin}), (\ref{tt19}) and (\ref{ttt26}), we rewrite $\bar{u}$ and $u_e$ as 
\begin{equation}\label{otrr11}
	\begin{aligned}
		\bar{u}
		=-k_nz_n-\frac{\hat{\delta}_{\theta}}{2}\left(|W_n|^2_\mathsf{F}+1\right)z_n-\frac{1}{2\epsilon_\Omega}z_n-\frac{1}{2}\epsilon_\Omega|\bar{\Omega}|^2 z_n\triangleq-\mathcal{K}(x,\hat{\theta},\hat{\delta}_{\theta})z_n,
	\end{aligned}
\end{equation}
\begin{equation}\label{otrr113}
	\begin{aligned}
		u_e
		=-\mathcal{K}(x,\hat{\theta},\hat{\delta}_{\theta})\hat{\rho}z_n,
	\end{aligned}
\end{equation}
where $\mathcal{K}(x,\hat{\theta},\hat{\delta}_{\theta})=k_n+\frac{1}{2}\left(\hat{\delta}_{\theta}|W_n|^2_\mathsf{F}+\hat{\delta}_{\theta}+\epsilon_\Omega|\bar{\Omega}|^2+\frac{1}{\epsilon_\Omega}\right)>0$ due to its construction.
 
By substituting (\ref{otrr11}) into (\ref{tt12}), yields $\dot{\hat{\rho}}=\gamma_\rho\sign{\ell_b}\mathcal{K}z^2_n$. Now, let us consider two cases for discussion.
\begin{itemize}
	\item When $b(t)>0$, for all $t\textgreater0$, Assumption \ref{assumption2} ensures the existence of a constant $\ell_b$ that satisfies $0<\ell_b<b(t)$, $~\Delta_b>0$, and $\dot{\hat{\rho}}(t)\textgreater0$. This indicates that an initialization with $\hat{\rho}(0)\geq0$ guarantees that $\hat{\rho}(t)>0$ for all  $t\textgreater0$, and hence  $-\Delta_b\hat{\rho}\mathcal{K}z^2_n\leq0$ for all $t\textgreater0$.
	\item Similarly, when $b(t)<0$ for all $t\textgreater0$, there exists a constant $\ell_b$ such that $b(t)\leq\ell_b<0, ~\Delta_b<0$, and $\dot{\hat{\rho}}(t)\leq0$. Choosing $\hat{\rho}(0)<0$ guarantees that $\hat{\rho}(t)<0$ for all $t\textgreater0$, and hence $-\Delta_b\hat{\rho}\mathcal{K}z^2_n\leq0$.
\end{itemize}
In both cases, $-\Delta_b\hat{\rho}\mathcal{K}z^2_n\leq0$ is guaranteed by choosing suitable initial values of $\hat\rho(0)$.

Finally, by recalling (\ref{ttt27}), and invoking $-\Delta_b\hat{\rho}\mathcal{K}z^2_n\leq0$, we can obtain:
\begin{equation}\label{otr13}
	\begin{aligned}
		\dot{V}\leq&-\sum_{j=1}^{n}k_jz^2_j+|\bar{b}|\gamma_u\sigma\leq-kV_z+|\bar{b}|\gamma_u\sigma,
	\end{aligned}
\end{equation}
where $k=2\min\left\{k_1,k_2,\cdots,k_n\right\}$.

	\begin{remark}
		The term ``tuning function'' was introduced in \cite{1995Krstic} and has been widely utilized in subsequent studies, such as \cite{WXY2001}. As these references have pointed out, the use of tuning functions can circumvent the inherent over-parametrization of traditional adaptive backstepping control methods. By the way, according to reference \cite{1995Krstic}, ``overparameterization'' refers to the necessity of continuously updating as many as $nq$ estimates of $q$ unknown parameters ($n$ represents the system order). In our work, overparameterization is avoided by introducing tuning functions that require a two-level adaptive update process and it can be checked that only $q+1$ estimations are needed for $q$ unknown time-varying parameters. 
\end{remark}

\section{Stability analysis}\label{sec-4}
In this section, we state the main result and give the stability analysis.

\begin{theorem}\label{theorem-1}
Suppose that Assumptions \ref{assumption1} and \ref{assumption2} hold. Consider a closed-loop system consisting of system (\ref{system1}), the event-triggering mechanism (\ref{tt3})-(\ref{tt4}), and the adaptive controller 
\begin{equation}
	\left\{\begin{array}{ll}  
			\dot{\hat{\theta}}=\Gamma\sum_{i=1}^{n}\omega_iz_i,\\
			\dot{\hat{\rho}}=-\gamma_\rho\sign{\ell_b}\bar{u}z_n,\\
			\dot{\hat{\delta}}_{\theta}	=\gamma_\delta\sum_{i=1}^{n}\frac{1}{2}\left(|W_i|^2_\mathsf{F}+n+1-i\right)z^2_i,\\
			\bar{u}		=-k_nz_n-\frac{\hat{\delta}_{\theta}}{2}\left(|W_n|^2_\mathsf{F}+1\right)z_n-\frac{1}{2\epsilon_\Omega}z_n-\frac{1}{2}\epsilon_\Omega|\bar{\Omega}|^2 z_n,
	\end{array}\right.
\end{equation} 
then the closed-loop system is globally uniformly asymptotically stable in the sense that all internal signals are bounded and ultimately converge to zero. In addition, the Zeno behavior of the triggering sequence $(t_k,u_e(t_k))_{k\in\mathbb{N}}$ is avoided.
\end{theorem}
	\textit{Proof: } Firstly, recalling (\ref{tt4}), (\ref{bt_1}), and (\ref{Omega_defin}), we have $|u_e(t)-u(t)|\leq\gamma_u$ and
\begin{equation}\label{ottr14}
	\int_{0}^{\infty}|\bar{b}|\gamma_u\sigma dt =\int_{0}^{\infty}|\bar{b}|\gamma_u e^{-\xi t}dt=\frac{|\bar{b}|\gamma_u}{\xi}<\infty,
\end{equation}
by taking the integration of (\ref{otr13}) over the interval $[0,t)$ yields
\begin{equation}\label{ottr2}
	V(t)\leq V(0)-\int_{0}^{t}kV_z dv+\int_{0}^{t}|\bar{b}|\gamma_u\sigma dv<\infty,
\end{equation}
then we can go further
\begin{equation}\label{ottr3}
	\int_{0}^{t}kV_z dv\leq V(0)+\int_{0}^{t}|\bar{b}|\gamma_u\sigma dv<\infty.
\end{equation}
It is known from (\ref{ottr2}) that $V(t)$ is bounded for $\forall t\in[0,\infty)$, which ensures that $V_z,~V_\theta,~V_\rho$, and $V_\delta$ are bounded, and we can infer that $z_i~(i=1,2,\cdots,n)$, $\hat{\theta}$, $\hat{\rho}$ and $\hat{\delta}_\theta$ are bounded as well. In addition, according to (\ref{ottr3}), $z_i\in\mathcal{L}_\infty\cap\mathcal{L}_2$, thus $\lim_{t\rightarrow \infty}z_i=0$ by applying Barbalat's lemma. From (\ref{z_1}), it follows that $\lim_{t\rightarrow \infty}x_1=0$. With definition of $\alpha_{1}$ given in (\ref{alpha_1}), it can be verified that $\lim_{t\rightarrow \infty}\alpha_1=0$. Furthermore, due to $x_2=z_2+\alpha_{1}$, it follows that $\lim_{t\rightarrow \infty}x_2=0$. Similarly, we can deduce that $\lim_{t\rightarrow \infty}\alpha_i=0$ and $\lim_{t\rightarrow \infty}x_i=0$. From (\ref{otrr11}), we know that the actual control input $\bar{u}$ is bounded, and $\lim_{t\rightarrow \infty}\bar{u}=0$. Since $\hat{\rho}\in\mathcal{L}_\infty$ and $u_e=\hat{\rho}\bar{u}$,  $u_e\in\mathcal{L}_\infty$ holds. Finally, the event-triggered mechanism (\ref{tt3}) guarantees that $u$ remains bounded at all times. Hence, the control objective is achieved.

Now we show that our proposed control protocol can avoid the
Zeno behavior, i.e. the phenomenon that the event is triggered for
infinite times in a finite time interval. To show this, we need to prove that
the inter-execution intervals are lower bounded by a positive constant. Here, we define $e(t)=u_e(t)-u(t)$. Therefore, recalling $u(t)=u_e(t), \forall \in [t_k, t_{k+1})$, which implies that
\begin{equation}\label{dot_e}
	\frac{d}{dt}|e|=\frac{d}{dt}(e*e)^{\frac{1}{2}}=\sign{u_e(t)-u(t)}\dot{u}_e\leq|\dot{u}_e|.
\end{equation}
From $u_e=\hat{\rho}\bar{u}$, it follows that $u_e$ is differentiable and $\dot{u}_e$ is a function of all bounded closed-loop signals. Therefore, there must exist a positive constant $\mu$ such that $\dot{u}_e\leq \mu$. This implies that $\frac{e(t_{k+1})-e(t_k)}{t_{k+1}-t_k}=\dot{e}\leq\mu$. Given that $e(t_k)=0$ and $\lim_{t\rightarrow t_{k+1}}e(t)=\gamma_u$, it can be inferred that the inter-execution intervals have a lower bound  $t_{k+1}-t_k\geq\frac{\gamma_u}{\mu}$, thereby successfully excluding the Zeno behavior. This completes the proof of Theorem \ref{theorem-1}. 
 $\hfill\blacksquare $
 
 \begin{remark}
 	From (\ref{tt2}), it can be deduced that the adaptive parameter $\hat{\theta}$ may drift if $z_1,\dots,z_n$ do not converge to zero. However, we have proved that $z_1,\cdots,z_n$ are bounded and ultimately converge to zero by using the fact $z_i\in\mathcal{L}_{\infty}\cap\mathcal{L}_2$, and thus the adaptive parameter $\hat{\theta}(t)$ will converge to a constant.
 \end{remark}
\begin{remark}
Our analysis is partly motivated by \textit{Chen et al.}\cite{Ye2022,Chen2018}, which presented \textit{congelation of variables} methods to deal with uncertain time-varying parameters. The difference is that this article integrates event-triggered mechanisms and establishes three separate adaptive laws (\ref{tt2}), (\ref{tt12}), and (\ref{tt21}) for controller design. 
\end{remark}
\section{Numerical example}\label{sec-5} 
In this section, we present a representative example to validate the effectiveness of the proposed control scheme.
Consider the following strict-feedback nonlinear system with time-varying parameters as
	\begin{equation}
		\left\{\begin{array}{ll}
		\begin{aligned}
			&\dot{x}_1=x_2+\theta(t) x^2_1,\\
			&\dot{x}_2=b(t)u,
		\end{aligned}
	\end{array}\right.
	\end{equation}
	with fast time-varying parameters
	\begin{equation}
		b(t)=2+0.1\cos(x_1)+0.5\sin(x_2)+0.5\operatorname{sgn}\left({x_1x_2}\right),
	\end{equation}
	\begin{equation}
		\theta(t)=2+0.8\sin (2t)+\sin(x_1x_2)+0.2\sin(x_1t)+\operatorname{sgn}\left({\sin(t)}\right).
	\end{equation}

It can be verified that both $b(t)$ and $\theta(t)$ satisfy Assumptions \ref{assumption1}--\ref{assumption2} with $\bar{b}=3.1$. To demonstrate the superiority of the proposed control method, we compare it with other two controllers. 
	\begin{itemize}
		\item Baseline, the classical adaptive backstepping controller proposed of \textit{Krstic et al.} \cite{1995Krstic};
		\item Controller 1, the event-triggered adaptive controller presented in \cite{XLT20172};
		\item and Controller 2 is the controller introduced in this article, described by equations (\ref{tt3})-(\ref{tt4}), (\ref{tt2}), (\ref{tt12}), (\ref{tt21}), and (\ref{otrr11}).
	\end{itemize}   
For the Baseline controller \cite{1995Krstic}, its control scheme is as follows:
\begin{equation}
	\begin{aligned}
		\left\{\begin{array}{ll}
			&z_1=x_1,\quad z_2=x_2-\alpha_{1},\\
			&\alpha_{1}=-k_1x_1-x^2_1\hat{\theta},\\
			&u=\frac{1}{b}\left(-k_2z_2-z_1-z_2\frac{\partial \alpha_{1}}{\partial x_1}-\alpha_{1}\frac{\partial \alpha_{1}}{\partial x_1}-x^2_1\hat{\theta}\frac{\partial \alpha_{1}}{\partial x_1}-\frac{\partial \alpha_{1}}{\partial \hat{\theta}}\dot{\hat{\theta}}\right),\\
			&\dot{\hat{\theta}}=\Gamma z_1x^2_1-\Gamma x^2_1\frac{\partial \alpha_{1}}{\partial x_1}.
		\end{array}\right.
	\end{aligned}
\end{equation}
For the Controller 1 \cite{XLT20172}, its control scheme is as follows:
\begin{equation}
	\begin{aligned}
		\left\{\begin{array}{ll}
			&z_1=x_1,\quad z_2=x_2-\alpha_{1},\\
			&\alpha_{1}=-k_1x_1-x^2_1\hat{\theta},\\
			&u=\frac{1}{b}\left(-k_2z_2-z_1-z_2\frac{\partial \alpha_{1}}{\partial x_1}-\alpha_{1}\frac{\partial \alpha_{1}}{\partial x_1}-x^2_1\hat{\theta}\frac{\partial \alpha_{1}}{\partial x_1}-\frac{\partial \alpha_{1}}{\partial \hat{\theta}}\dot{\hat{\theta}}-\bar{m}\tanh\left(\frac{z_2\bar{m}}{\epsilon}\right)\right),\\
			&\dot{\hat{\theta}}=\Gamma z_1x^2_1-\Gamma x^2_1\frac{\partial \alpha_{1}}{\partial x_1}-\sigma\hat{\theta}.
		\end{array}\right.
	\end{aligned}
\end{equation}
Based on the control scheme as given in Theorem \ref{theorem-1}, one gets the following control scheme for Controller 2: 
\begin{equation}\label{mechanmis-1}
	\begin{aligned}
		\left\{\begin{array}{ll}
	&u_e=-\hat{\rho}\mathcal{K} z_2,\\
	& \dot{\hat{\theta}}=\Gamma x^3_1+\Gamma z_2\frac{\partial \alpha_{1}}{\partial x_1}x^2_1,\quad \hat{\theta}(0)=0,\\
	&\dot{\hat{\rho}}=\gamma_\rho\operatorname{sgn}\left({\ell_b}\right)\mathcal{K} z^2_2,\quad \hat{\rho}(0)=0.4,\\
	&\dot{\hat{\delta}}_{\theta}
	=\frac{1}{2}\left(|W_2|^2_\mathsf{F}+1\right)z^2_2+\frac{1}{2}\left(x^{2}_1+2\right)x^2_1,\quad \hat{\delta}_{{\theta}}(0)=0,\\
		&\left\{\begin{array}{ll}
				&\mathcal{K}=k_2+\frac{1}{2}\left(\hat{\delta}_{\theta}|W_2|^2_\mathsf{F}+\hat{\delta}_{\theta}+\epsilon_\Omega|\bar{\Omega}|^2+\frac{1}{\epsilon_\Omega}\right),\quad
			\begin{array}{ll}
				W_2=\left[-\frac{\partial \alpha_{1}}{\partial x_1}x_1, ~0\right]^{\top},
			\end{array}\\
		&z_2 = x_2-\alpha_1,\quad \alpha_{1}=-k_1x_1-x^2_1\hat{\theta}-\frac{\hat{\delta}_{\theta}}{2}\left(x^{2}_1+2\right)x_1-\frac{1}{2\epsilon_\Omega}x_1,\\
		&\begin{array}{ll}
			\bar{\Omega} =\left[\begin{array}{c}
				1-\frac{\partial \alpha_{1}}{\partial x_1}\left(-k_1-\frac{\hat{\delta}_{\theta}}{2}\left(x^{2}_1+2\right)-\frac{1}{2\epsilon_\Omega}\right)-\frac{\partial \alpha_{1}}{\partial \hat{\theta}}\Gamma x^2_1-\frac{\gamma_\delta}{2}\frac{\partial \alpha_{1}}{\partial \hat{\theta}}\left(x^{2}_1+2\right)x_1\\
				\frac{\partial \alpha_{1}}{\partial x_1}\left(1+x^2_1\right)-\frac{\gamma_\delta}{2}\frac{\partial \alpha_{1}}{\partial \hat{\delta}_{{\theta}}}(|W_2|_{\mathsf{F}}^2+1)z_2+\frac{|\bar{b}|\gamma_u}{\sqrt{z^2_2+\sigma^2}}
			\end{array}\right].
		\end{array}
		\end{array}\right.
	\end{array}\right.
\end{aligned}
\end{equation}

To compare fairly, we initialize the system states as $[x_1(0);x_2(0)]=[1;-4]$, and select the following design parameters: $k_1=0.65$, $k_2=0.05$, $\Gamma=0.01$, and $\hat{\theta}(0)=0$. For Controller 1, we set the design parameters to $m=1$, $\bar{m}=2$, and $\epsilon=0.01$. For Controller 2, we set the parameters to $\epsilon_\Omega=5$, $\gamma_\delta=\gamma_\rho=0.01$, $m=1$ and $\gamma_u=0.1$.
	
\begin{figure}[!]
	\centering 
	\includegraphics[width=3in]{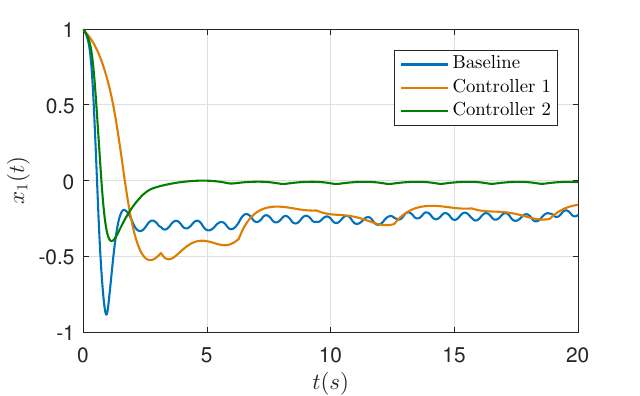}
	\includegraphics[width=3in]{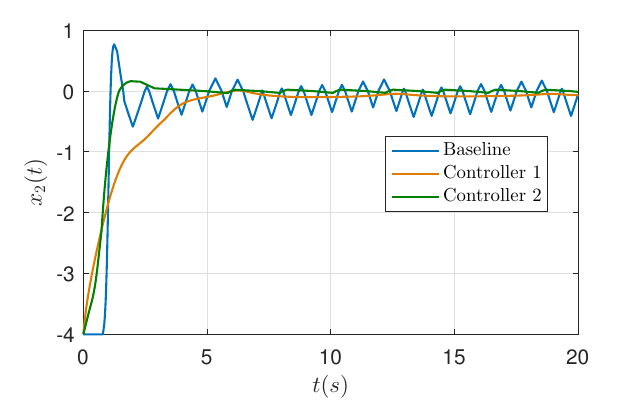}
	\caption{Trajectories of states $x_1(t)$ and $x_2(t)$ under different controllers.}	\label{fig-1}
\end{figure}

\begin{figure}[!]
	\centering 
	\includegraphics[width=5.3in]{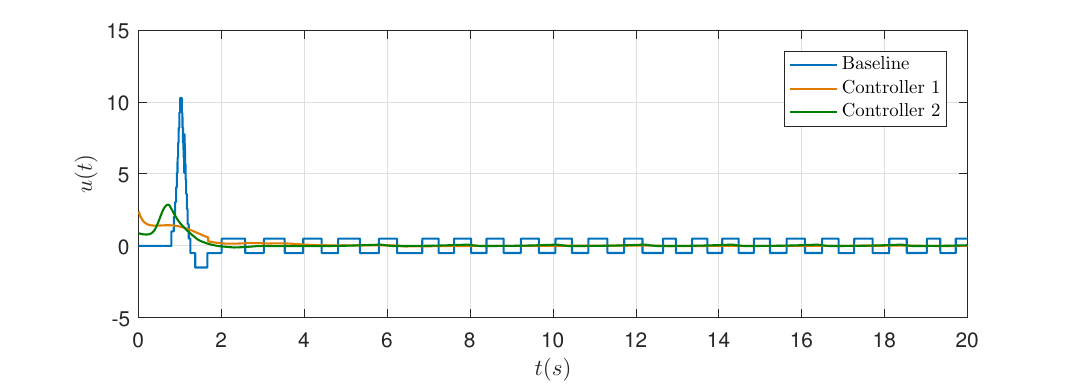}
	\caption{Trajectories of control input $u(t)$ under different controllers.}	\label{fig-2}
\end{figure}

\begin{figure}[!]
	\centering 
	\includegraphics[width=3in]{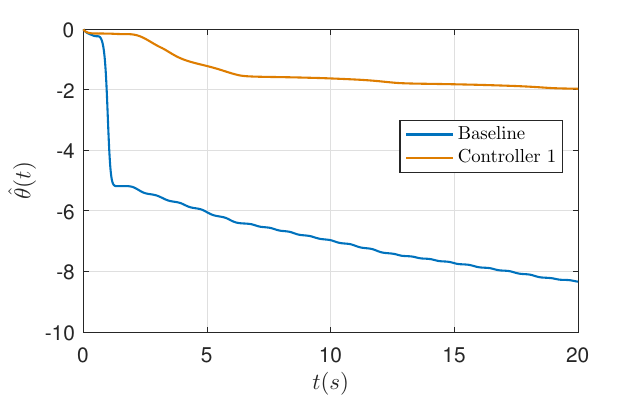}
	\includegraphics[width=3in]{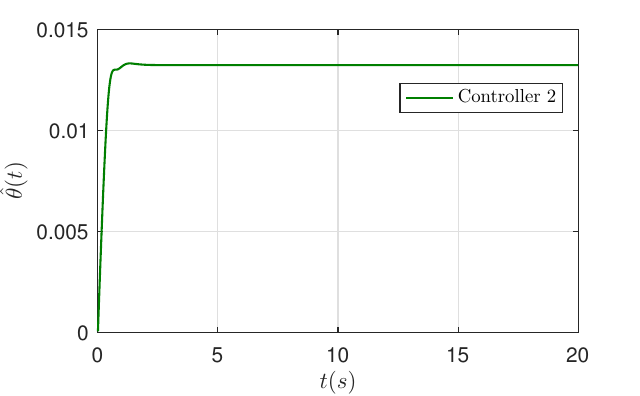}
	\caption{Trajectories of adaptive parameter $\hat{\theta}(t)$ under different controllers. }	\label{fig-3}
\end{figure}

\begin{figure}[!]
	\centering 
	\includegraphics[width=3in]{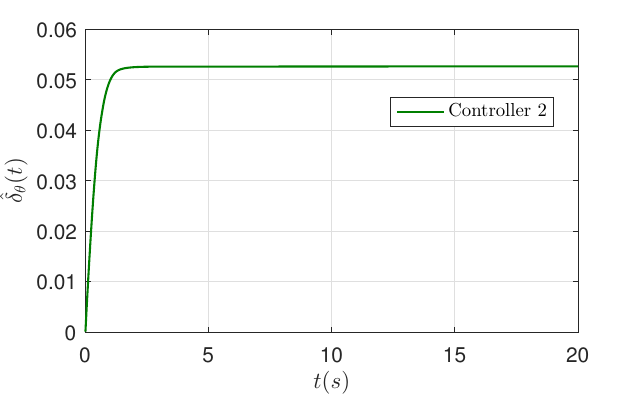}
	\includegraphics[width=3in]{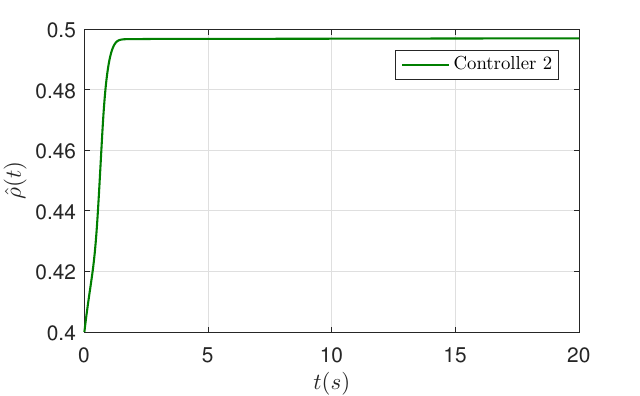}
	\caption{Trajectories of system time-varying parameters $\hat{\delta}_\theta(t)$ and $\hat{\rho}(t)$ under the proposed controller.}	\label{fig-5}
\end{figure}

\begin{figure}[!]
	\centering 
	\includegraphics[width=3in]{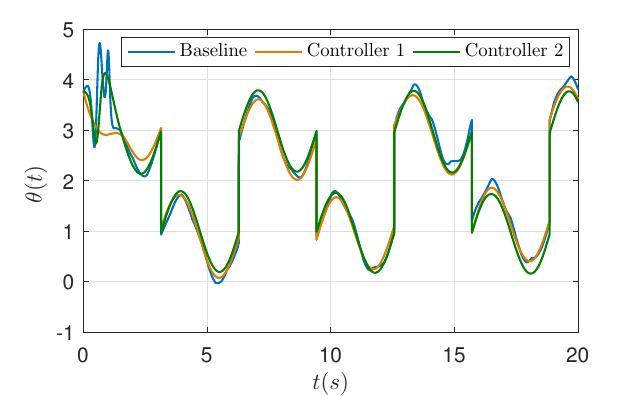}
	\includegraphics[width=3in]{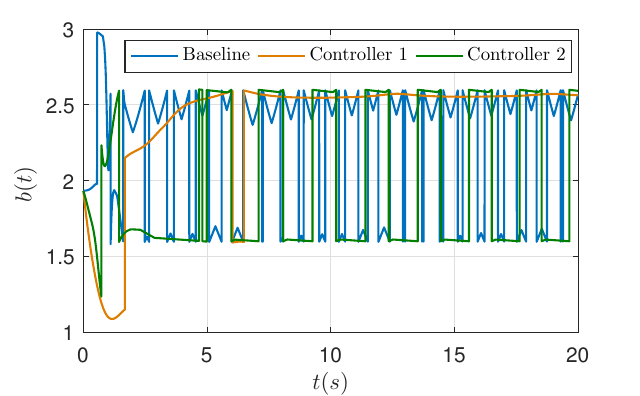}
	\caption{Trajectories of adaptive parameters $\theta(t)$ and $b(t)$ under different controllers.}	\label{fig-4}
\end{figure}

\begin{figure}[!]
	\centering 
	\includegraphics[width=3in]{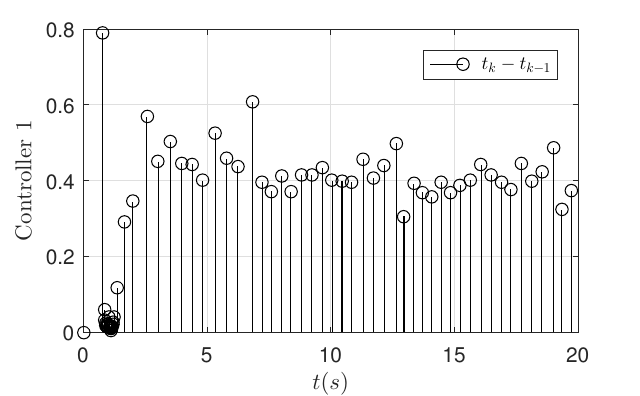}
	\includegraphics[width=3in]{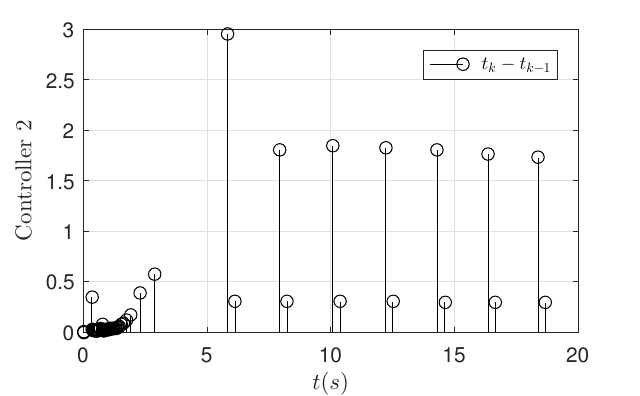}
	\caption{Trajectories of time interval of triggering events under the proposed controller.}	\label{fig-6}
\end{figure}

The simulation results, presented in Figs. \ref{fig-1}-\ref{fig-6}. Fig. \ref{fig-1} illustrates the trajectories of the state signals $x_1(t)$ and $x_2(t)$ under each controller. Fig. \ref{fig-2} reveals the evolution of the control input signals $u(t)$ under each controller. Fig. \ref{fig-3} displays the adaptive parameters $\hat{\theta}(t)$ under each controller. As shown in Fig. \ref{fig-3}, we observe that the estimated parameter $\hat{\theta}(t)$ ultimately converges to a constant value. In contrast, under the other two controllers, the estimated parameter $\hat{\theta}(t)$ does not converge to constant values (i.e., parameter drift occurs). In Fig. \ref{fig-5}, we present the progress of adaptive parameters $\hat{\delta}_\theta(t)$ and $\hat{\rho}(t)$ under Controller 2. The time-varying nature of parameters $\theta(t)$ and $b(t)$ under each controller is shown in Fig. \ref{fig-4}, highlighting that they may be non-smooth or may change suddenly over a period of time. Lastly, Fig. \ref{fig-6} visually represents the time intervals for each event. These results show that the proposed control algorithm outperforms \cite{XLT20172} and \cite{1995Krstic} since there is no parameter drift in this algorithm compared to the other two algorithms and all the signals of the closed-loop system eventually converge to zero, even when confronted with unknown and persistently varying parameters in the feedback and input paths. This outcome aligns with our theoretical findings. 

\section{Concluding remarks}\label{sec-6}
In this study, we develop a new event-triggered adaptive asymptotic control scheme for parameter-varying strict-feedback nonlinear systems. The basic design idea is derived from classical adaptive backstepping design, with the difference that a two-level adaptive parameter estimation strategy and two tuning functions are embedded in the design procedure, which has the advantage of relaxing the restrictive conditions in the existing literature that require the unknown parameters to be constant or bounded by a known constant. In addition, an event-triggering mechanism is designed to reduce data transmission, and the sampling error due to the triggering mechanism is handled by introducing a soft sign function to ensure that the estimated parameters are always bounded and do not drift. The Zeno behavior of the triggering sequence is also excluded through rigorous analysis. Simulation results verify the effectiveness of the proposed adaptive control scheme. In the future, it is of interest to robustify event-triggered control for measurement noise and actuator faults \cite{YHF20234,YHF20233}.  

\section*{Acknowledgments}
The authors would like to thank the editors and anonymous reviewers for their comments and constructive suggestions
that helped to improve this article significantly.

\section*{Conflict of interest}
All authors declare that they have no conflict of interests.

\section*{Data Availability statement}
Data sharing not applicable to this article as no datasets were generated or analyzed during the current study.


\begin{thebibliography}{99}
	\bibitem{2022-zhangzhirong-auto}
	Zhang ZR, Wen CY, Zhao K, and Song YD. Decentralized adaptive control of uncertain interconnected systems with triggering state signals, {\it Automatica,} 2022;141:110283.	
	\bibitem{ZQX2019}
	Zhu QX, Stabilization of stochastic nonlinear delay systems with exogenous disturbances and the event-triggered feedback control. {\it IEEE Trans. Automa. Control}, 2019;64(9):3764-3771.
	\bibitem{ZGP2021}
	Zhang GP and Zhu QX. Event-triggered optimized control for nonlinear delayed stochastic systems. {\it IEEE Trans. Circuits Syst. I-Regul. Pap.} 2021;68(9):3808-3821.
	\bibitem{2022-zhangzhirong-tac}
	Zhang ZR, Wen CY, Xing LT and Song YD. Adaptive event-triggered control of uncertain nonlinear systems using intermittent output only, {\it IEEE Trans. Automa. Control}, 2022;67(8):4218-4225.
	
	\bibitem{ZhangXM2023}
	Zhang XM, Han QL, Ge XH, Ning BD, Zhang BN. Sampled-data control systems with non-uniform sampling: A survey of methods and trends, {\it Annu. Rev. Control}, 2023; 55: 70-91.
	\bibitem{QWH2023}
	Lu LH, Xia MY, Qi WH, Yan HC, He HF, Cheng J. Adaptive event-triggered resilient stabilization for nonlinear semi-Markov jump systems subject to DoS attacks, {\it Int. J. Robust Nonlinear Control}. 2023;33(8):1914-1929.
	\bibitem{QWH2022}
	He HF, Qi WH, Cao JD, Cheng J, Shi KB. Observer-based adaptive sliding-mode control under dynamic event-triggered scheme with actuator and communication faults. {\it Int. J. Robust Nonlinear Control}. 2022; doi. https://doi.org/10.1002/rnc.6320.
	\bibitem{QWH2021}
	Qi WH, Hou YK, Zong GD and Ahn CK. Finite-time event-triggered control for semi-markovian switching cyber-physical systems with FDI attacks and applications, {\it IEEE Trans. Circuits Syst. I-Regul. Pap.}, 2021;68(6):2665-2674.
	\bibitem{QWH2022-1}
	Qi WH, Yi YJ, Zong GD, Ahn KC and Yan HC. Finite-time control for discrete-time positive systems subject to event-triggered scheme and markov jump parameters. {\it IEEE Trans. Circuits Syst. II-Express Briefs} 2022;69(12):4969-4973.
	\bibitem{WPMH2013}
	W. P. M. H. Heemels, M. C. F. Donkers and A. R. Teel. Periodic event-triggered control for linear systems. {\it IEEE Trans. Automa. Control.} 2013;58(4):847-861.
	\bibitem{WY2022}
	Wang Y, Wang XY, Zhao Y, Chen YQ, Li WQ. Finite-time event-triggered bumpless transfer control for switched systems. {\it Int. J. Robust Nonlinear Control}. 2022;32(12):6962-6982.
	\bibitem{XieYK2023}
		Xie YK, Ma Q and Xu SY. Adaptive event-triggered finite-time control for uncertain time delay nonlinear system. {\it IEEE T. Cybern.} 2023;53(9):5928-5937.
	\bibitem{Kaze2022}
		Kazemy A, Lam J and Zhang XM. Event-triggered output feedback synchronization of master–slave neural networks under deception attacks. {\it IEEE Trans. Neural Netw. Learn. Syst.} 2022;33(3):952-961.
	\bibitem{GeXH2022}
		Ge XH,  Xiao SY, Han QL, Zhang XM and Ding D. Dynamic event-triggered scheduling and platooning control co-design for automated vehicles over vehicular Ad-Hoc networks. {\it IEEE-CAA J. Automatica Sin.} 2022;9(1):31-46. 
	\bibitem{GuZ2023}
		Gu Z, Yue D, Ahn CK, Yan S and Xie XP. Segment-weighted information-based event-triggered mechanism for networked control systems. {\it IEEE T. Cybern.} 2023;53(8):5336-5345.
	\bibitem{LiKun2023}
	Li K, Zhao K, Jia X, Gao R. Adaptive global prescribed performance tracking control of uncertain robotic systems with unknown control directions. {\it Int. J. Robust Nonlinear Control}. 2023;33(11):5956-5974. 
	\bibitem{Yantan2023}
	Tan Y, Wu LC. Neuro-adaptive practical prescribed-time control for pure-feedback nonlinear systems without accurate initial errors. {\it Int J Adapt Control Signal Process.} 2023;37(4):915-933.
	\bibitem{ZhaoK2016}
	Song YD, Zhao K and Krstic M, Adaptive control with exponential regulation in the absence of persistent excitation. {\it IEEE Trans. Automa. Control.} 2017; 62(5):2589-2596.
	\bibitem{ZhaoK2022}
	Zhao K, Song YD, Chen C. L. P and Chen L. Adaptive asymptotic tracking with global performance for nonlinear systems With unknown control directions. {\it IEEE Trans. Automa. Control.}  2022;67(3):1566-1573.
	\bibitem{Xing2017}
	Xing LT, Wen CY, Liu ZT, Su HY and Cai JP. Event-triggered adaptive control for a class of uncertain nonlinear systems. {\it IEEE Trans. Automa. Control.} 2017;62(4):2071–2076.
	\bibitem{XLT20174}
	Xing LT, Wen CY, Liu ZT, Su HY and Cai JP. Adaptive compensation for actuator failures with event-triggered input. {\it Automatica,} 2017;85:129-136.
	\bibitem{HuangYX2019}
	Huang YX and Liu YG. Practical tracking via adaptive event-triggered feedback for uncertain nonlinear systems. {\it IEEE Trans. Automa. Control.} 2019;64(9):3920-3927.
	\bibitem{LongJ2022}
	Long J, Wang W, Huang JS, L{\"u} JH and Liu KX. Adaptive leaderless consensus for uncertain high-order nonlinear multiagent systems with event-triggered communication. {\it IEEE Trans. Syst. Man Cybern. -Syst.} 2022;52(11):7101-7111.
	\bibitem{XLT20172}
	Xing LT, Wen CY, Liu ZT, Su HY and Cai JP. Event-triggered adaptive control for a class of uncertain nonlinear systems. {\it IEEE Trans. Automa. Control.} 2017;62(4):2071-2076.
	\bibitem{LD2023}
	Liu D, Liu N, Li TS. Event-triggered model-free adaptive control for nonlinear systems with output saturation. {\it Int. J. Robust Nonlinear Control}. 2023;https://doi.org/10.1002/rnc.6747.
	\bibitem{XLT20173}
	Xing LT, Wen CY, Liu ZT, Su HY and Cai JP. Adaptive compensation for actuator failures with event-triggered input. {\it Automatica,} 2017;85:129-136.
	\bibitem{KarafyllisI2018}
	Karafyllis I and Krstic M. Adaptive certainty-equivalence control with regulation-triggered finite-time least-squares identification. {\it IEEE Trans. Automa. Control.} 2018; 63(10):3261-3275.
	\bibitem{Huang2020}
	Huang JS, Wang W, Wen CY, and Li GQ. Adaptive event-triggered
	control of nonlinear systems with controller and parameter estimator
	triggering. {\it IEEE Trans. Autom. Control,} 2020;65(1):318–324.
	\bibitem{LiuCS2017}
	Liu CS, Ye Q, Zhang SJ. Adaptive control for a class of uncertain linear parameter-varying flight aircraft systems. {\it Int. J. Adapt. Control Signal Process.} 2017;31(2):210-222.
	\bibitem{OHSY2020}
	Oh S-Y, Choi H-L. Robust control for nonlinear systems with uncertain time-varying parameters coupled with non-triangular terms. {\it Int
	J Syst Sci.} 2020;51(3):507-521.
	\bibitem{YeHF20232}
	Ye HF, Song YD. Prescribed-time control for time-varying nonlinear systems: A temporal scaling based robust adaptive approach. {\it Syst. Control Lett.} 2023;181:105602.
	
	\bibitem{1995Krstic}
	Krstic M, Kanellakopoulos I, Kokotovic P.  Nonlinear and adaptive control design. {\it John Wiley and Sons,} 1995.
	\bibitem{Khalil2002}
	H. Khalil, Nonlinear Systems, {\it Englewood Cliffs, NJ, USA: Prentice Hall.} 2002.
	
	\bibitem{WCY2000}
	Zhang Y, Wen CY, Yeng Chai Soh. Discrete-time robust backstepping adaptive control for nonlinear time-varying systems. {\it IEEE Trans. Automa. Control.} 2000; 45(9):1749-1755.
	
	\bibitem{WXY2001}
	Zhang Y, Wen CY, Yeng Chai Soh. Robust adaptive control of nonlinear discrete-time systems by backstepping without overparameterization. 
	{\it Automatica,} 2001;37(4):551-558.
	\bibitem{Ye2022}
	Ye HF, Song YD. Adaptive control with guaranteed transient behavior and zero steady-state error for systems with time-varying parameters. {\it IEEE-CAA J. Automatica Sin.} 2022;9(6):1073–1082.
	
	\bibitem{Astolfi2021}
	Chen KW, Astolfi A. Adaptive control for systems with time-varying parameters. {\it IEEE Trans. Automa. Control.} 2021;66(5):1986–2001.
	
	\bibitem{DingZ2000}
	Ding, ZT. Adaptive control of nonlinear systems with unknown virtual control coefficients. {\it Int J Adapt Control Signal Process. } 2000;14(5):505–517.

	\bibitem{YeHF2022}
	Ye HF, Zhao K, Wu HJ and Song YD. Adaptive control with global exponential stability for parameter-varying nonlinear systems under unknown control gains. {\it IEEE Trans. Cyberne}. 2023; 52(12),7858-7867.
	\bibitem{YeHF20222}
	Ye HF, Song YD. Backstepping design embedded with time-varying command filters. {\it IEEE Trans. Circuits Syst. II-Express Briefs}, 2022; 69(6):2832-2836.
	\bibitem{Nestruev2006}
	J. Nestruev. Smooth manifolds and observables. {\it Berlin, Germany: Springer,} 2006.
	
	\bibitem{Chen2018}
	Chen KW, Astolfi A. Adaptive control of linear systems with time-varying parameters. {\it Proc. Annu. Amer. Control Conf}, 2018; 80–85.
	\bibitem{YHF20234}
	Ye HF, Polycarpou MM, \& Wen CY. Matrix pencil based robust control for feedforward systems with event-triggered communications and sensor/actuator faults. {\it J. Automat. Intell.} 2023; 2(3), 139-145.
	\bibitem{YHF20233}
	Ye HF, Song YD, Zhang ZR, \& Wen CY. Global dynamic event-triggered control for nonlinear systems with sensor and actuator faults: A matrix pencil based approach. {\it IEEE Trans. Automa. Control.} doi: 10.1109/TAC.2023.3313634.
\end{thebibliography}
\end{document}